\documentstyle[12pt]{article}
\newcommand{\sg}{\sigma_{\pi N}}
\newcommand{\Sg}{\Sigma_{\pi N}}
\renewcommand{\ss}{\langle N|\bar{s}s|N\rangle}
\newcommand{\ud}{\langle N|\bar{u}u + \bar{d}d |N\rangle}
\newcommand{\uu}{\langle p|\bar{u}u |p\rangle}
\newcommand{\dd}{\langle p|\bar{d}d |p\rangle}
 
\begin{document}
 
\begin{titlepage}
\begin{flushright}
       {\bf UK/95-12}  \\
 Dec. 1995   \\
       hep-ph/9602259 \\
\end{flushright}
\begin{center}
 
{\bf {\LARGE $\pi N \sigma$ Term , $\bar{s}s$ in Nucleon, and
Scalar Form Factor --- a Lattice Study}}
 
\vspace{1cm}
 
{\bf S.J. Dong, J.-F. Laga\"{e}{\footnote
{Present Address: High Energy Physics Division, Argonne National Laboratory,
Argonne, IL 60439}}, and K.F. Liu   } \\[0.5em]
 {\it  Dept. of Physics and Astronomy  \\
  Univ. of Kentucky, Lexington, KY 40506}
 
\end{center}
 
\vspace{0.4cm}
 
\begin{abstract}

We report on a lattice QCD
calculation of the $\pi N \sigma$ term, the scalar form factor, and
$\langle N|\bar{s}s|N\rangle$.
The disconnected insertion part of $\sg$ is found to be
$1.8 \pm 0.1$ times larger than the connected insertion contribution.
The $q^2$ dependence of $\sg(q^2)$ is about the same as $G_E(q^2)$ of
the proton so that $\sg(2m_{\pi}^2) - \sg(0) = 6.6 \pm 0.6$  MeV. The
ratio $y= \ss/\ud = 0.36 \pm 0.03$. Both results favor a  $\sg
\sim 53$ MeV, slightly larger than our direct calculation of
$\sg = 49.7 \pm 2.6$ MeV. We also compute $F_s$ and $D_s$ and find that
the agreement with those from the octect baryon mass splittings crucially
depends on the inclusion of the large disconnected insertion. Finally,
we give our result for the $K N \sigma$ term.

\bigskip
 
 
\end{abstract}
 
\vfill
 
\end{titlepage}
 
   Like the pion mass in the meson sector, the $\pi N \sigma$ term is
a measure of the explicit chiral symmetry breaking in the baryon
sector. It is considered a fundamental quantity
which pertains to a wide range of issues in the low-energy hadron physics,
such as quark and baryon masses, strangeness content of
the nucleon, pattern of SU(3) breaking, $\pi N$ and $K N$ scatterings,
kaon condensate in dense matter, trace anomaly, and decoupling of heavy
quarks. Defined as the double-commutator of the isovector axial
charge with the hamiltonian density taken between the nucleon states, i.e.
$\sigma_{\pi N} = 1/3 \sum_{a=1,3} \langle N| [Q_a^5, [Q_a^5, {
\cal H}(0)]] |N\rangle$
which appears in the off-shell $\pi N$ scattering amplitude \cite{wei66},
it has in QCD the expression
\begin{equation}
\sigma_{\pi N} =\hat{m}\langle N|\bar{u}u + \bar{d}d|N\rangle ,
\end{equation}
where $\hat{m} = (m_u + m_d)/2$.
 
It is shown \cite{cd71} that at lowest order in $m_{\pi}$
(i.e. $m_{\pi}^2$), it is equal to the unphysical, but on-shell,
isospin even $\pi N$ scattering amplitude
at the Cheng-Dashen point, $\Sigma_{\pi N}= f_{\pi}^2 \bar{T}^+
(s = m_N^2, t = q^2 = 2m_{\pi}^2)$. Thus $\Sigma_{\pi N}$ can be extracted
from $\pi N$ scattering experiment via fixed-t dispersion relation
for instance \cite{cd71}.
 It is further shown \cite{bpp71} that the next higher order term
 which is nonanalytic in quark mass (i.e. proportional to $\hat{m}^{3/2}$
 or $m_{\pi}^3$) drops out if $\Sigma_{\pi N}$ is identified with
 $\sg(2m_{\pi}^2)$ \cite{bpp71} which is only a function of $q^2$.
 This shows that the difference $\Delta_R$ in the relation
 $ \Sg = \sg(2m_{\pi}^2) + \Delta_R$
 is of the order $m_{\pi}^4/m_N^4$ and has been shown to be
 indeed negligible ($\sim$ 0.35 MeV) in a chiral perturbation calculation
 \cite{bpp71,gls91}.
 
 Various estimates of $\Sg$ have ranged from 22 to 110 MeV over the years,
 but eventually settled around 60 MeV \cite{gls91}.
 On the other hand, a puzzle was raised by Cheng
 \cite{che76}. If one assumes that $\langle N|\bar{s}s|N\rangle = 0$,
 a reasonable assumption from the OZI rule, the $\sigma_{\pi N}^{(0)}$
 obtained from the octet baryon masses gives only 32 MeV, almost
 a factor two smaller than $\Sg$ extracted from the $\pi N$ scattering.
 This puzzle was tackled from both ends. First, the scalar
 form factor was
 calculated \cite{gls91} in chiral perturbation theory ($\chi$ PT) with the
 two correlated pions as the dominating intermediate state.
 As a result, the scalar form factor is found to be exceedingly soft
 which leads to a large change of $\sg(q^2)$ in a small range of $q^2$,
 i.e.
  $\Delta \sigma_{\pi N} = \sigma_{\pi N}(2 m_{\pi}^2) - \sigma_{\pi N}
 (0) = 15.2 \pm 0.4$ MeV. Thus, this reduces $\sigma_{\pi N}$
 \begin{equation}
 \sigma_{\pi N} = \Sigma_{\pi N} - \Delta \sigma_{\pi N}
 \end{equation}
  to $\sim 45$ MeV. The remaining
 discrepancy between $\sigma_{\pi N}$ and $\sigma_{\pi N}^{(0)}$
 is reconciled if one admits the possibility
 of a large $\bar{s}s$ content in the nucleon
 \cite{che76,gl82}. From the pattern of SU(3) breaking in
 the octet baryon masses, one finds \cite{gl82,che76}
  \begin{equation}   \label{y}
 \sg = \sg^{(0)}/(1 - y),
 \end{equation}
 where $y= 2 \ss/\ud$. Given $\sg^{(0)} = 32 $MeV from the octet
 baryon masses \cite{che76}, or 35(5) MeV from
 the one loop $\chi$ PT calculation \cite{gl82} and $\sg =
 45$MeV, eq. (\ref{y}) implies $ y = 0.2 $--- 0.3.
 
 Hence, a consistent solution seems to have emerged which suggests that
$\sigma_{\pi N} \sim 45\, $MeV, $\Delta \sg \sim 15\,$ MeV, and
$y \sim 0.2$ --- 0.3. In this letter, we undertake a lattice QCD
calculation
of the above quantities to scrutinize the viability of this resolution.
It turns out that our study strongly points to a significantly
different solution as we shall show.

    The calculation of $\sg$ in lattice QCD has been attempted by
several groups \cite{gss88,gbb91}
who employed the Feynman-Hellman theorem
 
\begin{equation} \label{fht}
\hat{m} \frac{\partial M_N}{\partial \hat{m}} =
\hat{m}\langle N|\bar{u}u + \bar{d}d|N\rangle_{C.I.} +
\hat{m}\langle N|\bar{u}u + \bar{d}d|N\rangle_{D.I}
\end{equation}
and obtained $\sg$ through the derivative of the nucleon mass.
We note that in eq. (\ref{fht}) the {\it connected insertion
(C.I.)} part comes from the differentiation with respect to
the valence
quark propagator; whereas the {\it disconnected insertion (D.I.)}
comes from the derivative of the fermion determinant.
Their contributions to the scalar current
$\bar{\psi}\psi$ in the nucleon are shown schematically in Fig. 1.
In the quenched approximation approach, it is found that
$\sg$ obtained from the derivative of the nucleon mass
 is only about 15 --- 25 MeV \cite{gss88}. This is much smaller than
the phenomenological value of $\sim 45$ MeV \cite{gls91}.
 The smallness of $\sg$ in this case is traced to the fact that
the nucleon mass in the quenched approximation is calculated with
the determinant set to a constant, so that its derivative corresponds
to the
{\it C.I.} only (which is verifiable by comparing to the direct evaluation
of the {\it C.I.} \cite{mmp87,dl92}) and it does not involve {\it D.I.}
which can be substantial. Indeed, when the derivative of $M_N$ is
calculated with
dynamical fermions included, it is found \cite{gbb91} that the l.h.s. of
eq. (\ref{fht}) which now includes the {\it D.I.}
becomes $\sim 2$ to 3 times larger than the {\it C.I.} contribution. This
implies a large contribution of the {\it D.I.}. Since the error
on $\partial M_N/\partial \hat{m}$ is quite large \cite{dl92,gbb91},
we decided to calculate the {\it D.I.} directly \cite{dl92} with the
help of the $Z_2$ noise \cite{dl94}.
Following our calculation of
the flavor-singlet $g_A^0$ \cite{dll95}, we calculate the {\it C.I.} and
{\it D.I.} of $\sg$ directly in the quenched approximation.
In terms of the Feynman-Hellman theorem, it would
correspond to calculating $\partial M_N/\partial \hat{m}$ by
taking the derivative of the determinant first before setting it
to a constant.
 
Lattice calculations of three-point functions have been used to
study the EM~\cite{dwl90}, axial (isovector)
\cite{ldd94a}, pseudoscalar($\pi NN$) \cite{ldd94b} form factors,
and the flavor-singlet $g_A^0$ \cite{dll95}. For the scalar
current,
%
%
%
we use $S(x) \!=\! 2\kappa /8 \kappa_c [\bar{u}u(x) + \bar{d}d(x)]$,
where we have implemented the mean-field improvement factor
 $8 \kappa_c$ to define the lattice operator \cite{lm93}.
 
The {\it C.I.} is calculated in the same way as the isovector
axial coupling $g_A^3$ \cite{ldd94a}.
Numerical details are given in Ref. \cite{ldd94a}. Like in Ref.
\cite{dll95}, the lattice renormalized
$g_{S, con}^L = \ud_{con}^L $ has been calculated for
$\kappa = 0.154, 0.152$, and 0.148,
corresponding to quark masses of about 120, 200, and 370 MeV
respectively (the scale $a^{-1} = 1.74(10)$GeV is set by the nucleon
mass), and is plotted in Fig. 2(a). The calculations were done on a
quenched $16^3 \times 24$ lattice at $\beta = 6.0$ with 24 gauge
configurations as in the previous cases \cite{ldd94a,dll95}.
Due to the fact that the quenched \mbox{$\chi$ PT} calculation
exhibits a leading non-analytic behavior of $m^{3/2}$ for the
nucleon mass \cite{ls94}, we extrapolate $g_{S, con}^L$
to the chiral limit ($\kappa_c = 0.1568$) with the form $C + D m^{1/2}$.
This is so because $g_{S, con}^L = \partial M_N/\partial \hat{m}$
in the quenched approximation as we alluded to earlier in eq. (\ref{fht}).
As a result, we obtain $g_{S, con}^L
= 3.04(9)$ as shown in Fig. 2(a). The $g_S $ in the continuum
with the $\overline{MS}$ scheme
is related to its lattice counterpart by the relation $g_S
= Z_S g_S^L$, where $Z_S$ is the finite lattice renormalization
constant. The one-loop calculation gives $Z_S = 0.995$ for
$\beta = 6.0$~\cite{lm93}, from which we find
$g_{S, con} = 3.02 \pm 0.09$. We also computed  isovector $g_S^{3}
= \langle N|\bar{u}u - \bar{d}d|N\rangle$ which does not involve the
{\it D.I.} and find it to be $0.63(7)$.
 
Since $\sg$ is renormalization group invariant, the {\it C.I.}
contribution is  $\sigma_{\pi N, con} =
\hat{m} g_{S, con}^L$ where $\hat{m}$ is the lattice quark
mass. From $m_{\pi}^2$ and $M_N$, we find $\hat{m} = 5.84(13)$ MeV.
Thus, $\sigma_{\pi N, con} = 17.8 (9)$ MeV which agrees well with
previous calculations \cite{gss88,mmp87,fko95}.
The {\it C.I.} part of the form factor is obtained by extrapolating
$g_{S,con}^L(q^2)$ at different $\kappa$ to the chiral limit. It is
plotted in Fig. 2(b) together with $g_A^3(q^2)$, the isovector axial
form factor. We see that they are almost identical within errors. In so
far as the concept of meson dominance goes, this reflects
in part the fact that the isovector scalar meson and $A_1$ are
essentially degenerate in the lattice calculation.
Like in the case of the axial coupling constants \cite{dll95}, we
also find that the ratio $R_S = g_S^3/g_{S, con}$ dips below the SU(6)
result of 1/3 as the quark mass becomes lighter. This is interpreted
as due to the cloud quark/anti-quark effect and is responsible for
the $\bar{u}$ -- $\bar{d}$ parton difference reflected in the
Gottfried sum rule \cite{ld94}. Only when the cloud degree of freedom
is eliminated in the valence approximation \cite{ld94}
where the Fock space is limited to the valence do we recover
the SU(6) limit. This indirectly shows the effect of the cloud quarks
in the {\it C.I.}.
 
   We calculate the {\it D.I.} in Fig. 1(b) the same way we did
for the {\it D.I.} part of $g_A^0$ \cite{dll95} by summing
over t. For $t_f >> a$, this sum becomes
\begin{equation} \label{ratiodis}
\sum_t\frac{G_{PSP}^{\alpha\alpha}(t_f, \vec{p}, t, \vec{q})
G_{PP}^{\alpha\alpha} (t, \vec{p})}
 {G_{PP}^{\alpha\alpha}(t_f, \vec{p}) G_{PP}^{\alpha\alpha}(t, \vec{q})}
 \,\, {}_{\stackrel{\longrightarrow}{t_f >>a}}\,\, {\rm const}
 + t_f g_{S, dis}^{L}(q^2)
\end{equation}

Thus, we calculate the sum as a function of $t_f$ and
take the slope to obtain the {\it D.I.} part of $g_S^L$. Since
the {\it D.I.}
involves quark loops which entail the calculation of
traces of the inverse quark matrices, we use the proven
efficient algorithm
to estimate these traces stochastically with the $Z_2$ noise \cite{dl94}
which was applied to the study of $g_A^0$ \cite{dll95}.


     The results of eq. (\ref{ratiodis}) with 300 complex
$Z_2$ noise and 50 gauge configurations for $\kappa = 0.148,
0.152$ and 0.154 are presented in Fig. 3. The corresponding
$g_{S, dis}^L = \ud_{dis}^L$ are obtained from fitting the slopes
in the region $t_f \ge 8$ where the nucleon is isolated from its
excited states with the correlation among the time slices taken
into account \cite{dll95}.
 The resultant fits covering the ranges of
$t_f$ with the minimum $\chi^2$ are plotted in Fig. 3.
Finally, the errors on the fit, also shown in the figure, are
obtained by jackknifing the procedure.
 
Plotted in Fig. 4(a) are the results
of $g_{S, dis}^L$ with the same sea-quark mass as those of the
valence- (and cloud-) quarks in the nucleon.
They suggest
a non-linear behavior in the quark-mass. This is enhanced by our finding
of a very soft form factor (Fig. 4(b)) which is consistent with the
expectations of $\chi PT$ \cite{gl82}
where the pion loop leads to a non-analytic behavior in $m_q^{3/2}$.
Furthermore, this non-linear behavior is seen prominently in hadron
masses when
dynamical fermions are included \cite{ceh95}. For these reasons, we
fit $\ud_{dis}$ with the linear plus $m^{1/2}$ form as for the
{\it C.I.} and get a small $\chi^2$ (see Fig. 4(a)).
The extrapolation
to the chiral limit is carried out in the same way as in the
case of $g_A^0$ \cite{dll95}.
To calculate $\ss$, we fix the sea-quark mass
at 0.154 and extrapolate the valence-quark mass
to the chiral limit with the
form $C + D \sqrt{\hat{m} + m_s}$ to reflect the $m_K^3$ dependence
of the nucleon mass from the kaon loop in $\chi$ PT.
These results are also plotted in Fig. 4(a).
 
    From Fig. 4(a), we find that $\ud_{dis} = Z_S \ud_{dis}^L
 = 5.41(15)$. This is $1.8(1)$ times the {\it C. I.} and
is consistent with previous indirect calculations based on
$\partial M_N/\partial \hat{m}$ with dynamical fermions\cite{gbb91},
a direct calculation with staggered fermions \cite{agh94}, and
the recent direct calculation \cite{fko95}
which gives a ratio of $2.2(6)$.
Similarly, we find from Fig. 4(a)
that $\ss = Z_S f(ma) \ss^L = 1.53 (7)$ where we have included
the finite ma correction factor f(ma) = 0.79 which was computed by
comparing the triangle diagram in the continuum and on the lattice
\cite{ll95}. This is much smaller than the recent calculation
\cite{fko95} which gives $\ss = 2.84(44)$.
Part of the disagreement comes from
the fact that a finite ma correction factor which is only
appropriate for a {\it C. I.} was used in ref. \cite{fko95} for the
\mbox{{\it D.I.}}. This leads to an overestimate by $\sim 30\%$.
In addition, summing $\sum
_{\vec{x}} S(\vec{x},t)$ in eq. (\ref{ratiodis}) over the edges in
time where the fixed B. C. is applied as is done in
Ref. \cite{fko95} gives an unphysical effect.
Our difference might be reconciled by these two effects.
 
   From the above results, we list $\uu, \dd, \ss$, $F_S =
(\uu -\ss)/2$, and $D_S = (\uu - 2\dd + \ss)/2$ in
Table 1. We see that both $D_S$ and $F_S$ compare favorably with the
phenomenological values obtained from the SU(3) breaking pattern of
the octect baryon masses with either linear
\cite{mmp87,fko95} or quadratic mass relations \cite{gas81}.
Especially, we should point
out that the agreement is significantly improved from the valence
quark model which predicts $F_S < 1$ and $D_S = 0$ and also those of
the {\it C. I.} alone.
The latter yields $F_S = 0.91(13)$ and $D_S = - 0.28(10)$ which are
only half of the phenomenological values \cite{mmp87,fko95,gas81}.
This underscores
the importance of the sea-quark contributions. We also obtain the form
factor $g_{S,dis}^L(q^2) = \ud_{dis}^L(q^2)$ for the {\it D.I.}
as plotted in Fig. 4(b).
We see that it is exceedingly soft which is reminiscent of the
two $\pi$ intermediate state in the $\chi$ PT calculation
\cite{gls91}. This possibility can be seen in Fig. 1(b) with
two $\pi$ dominance.
Indeed, if we assume that the {\it D. I.} part completely saturates
$\sg$ with $g_S = 8.43(24)$, it would give $\Delta \sg
= 11.5(2.1) $ MeV similar to that of the $\chi PT$ calculation
\cite{gls91}. However, there is also the {\it C. I.} part (fig. 2(b))
which is much harder than the {\it D.I.} When combined, it yields
a scalar form factor $g_S(q^2)$ which is softer than $g_A^3(q^2)$
and becomes close to $G_E(q^2)$ of the proton. They are plotted in Fig. 5
for comparison. Fitting the
$g_S(q^2)$ to a dipole form gives a dipole mass $m_D = 0.80(4)$ GeV. This
predicts $ \Delta \sg = 6.6(6)$ MeV, much smaller than the 15.2(4) MeV
obtained solely based on the two-$\pi$ dominance. We conclude from
this that the $\chi PT$ calculation \cite{gls91} is relevant to the
{\it D.I.} but missed the {\it C.I.} which maybe dominanted by
the scalar meson. On the other hand, the $\ss (q^2)$ comes only from
the {\it D.I.}, hence is very soft. Its r.m.s. radius
$\langle r^2\rangle_S^{1/2}(\bar{s}s) = 1.06(9)$ fm can be
interpreted as the size of the $K\bar{K}$ meson cloud in the scalar
channel (see Fig. 1(b)).
 
 For the parameter $y$ in eq. (\ref{y}), we find it to be 0.36(3).
Both $\Delta \sg$ and $y$ differ
significantly from the phenomenological solution based on $\chi PT$
as mentioned earlier which did not take into account the {\it C. I.}
with a possible scalar dominance. Our results on $\Delta \sg$ and
$y$ strongly suggest a higher $\sg = \Sg - \Delta \sg \sim
53$ MeV, assuming $\Sg \sim 60$ MeV and $\sigma^{(0)} \sim 32 $ ---
35 MeV.
Now, we turn to our result of $\sg$. Our direct calculation gives
$\ud = 8.43(24) $  and $\sg = 49.7(2.6)$ MeV. This is about one and
half $\sigma$ smaller than 53 MeV inferred from $\Delta \sg$ and $y$.
Since the direct computation of
$\sg$ involves the determination of the quark mass which is more
susceptible to systematic errors (such as the extrapolation in the
quark mass and the infinite volume limit) than the $q^2 $
dependence of the form factor and the ratio $y$, we believe
that our result on $\sg$ is less reliable than $\Delta \sg$ and
$y$. To examine the sensitivity of these three quantities as far as
the chiral limit extrapolation is concerned, we fit them to a linear
function in $m$ instead of $m^{1/2}$ and find that $\Delta \sg
= 4.7(8)$ MeV, $ y= 0.42(3)$, and $\sg = 39.0(2.0)$ MeV. Again, we
see that both $\Delta \sg$ and $y$ favor a higher $\sg \sim 55$ MeV which
is very close to the above estimate of 53 MeV with the $m^{1/2}$
extrapolation. Yet, the directly calculated $\sg$ falls short of this
expectation and is also much smaller than that of the $m^{1/2}$
extrapolation.

 \begin{table}[ht]
\caption{Scalar contents, $\Delta \sg$, y and $\sg$ in
comparison with phenomenology}
\begin{tabular}{llll}
 \multicolumn{1}{c}{} &\multicolumn{1}{c}{Lattice}
 & \multicolumn{1}{c} {Phenomenology} \\
 \hline
 $\uu$   & 4.55(16) &   \\
 $\dd$ & 3.92(16) & \\
 $\ss$  & 1.53(7) &   \\
 $F_S$ & 1.51(12) & 1.52 \cite{mmp87} --- 1.81\cite{gas81} \\
 $D_S$ &  -0.88(28)& -0.52\cite{mmp87} --- -0.57\cite{gas81}  \\
 $\langle r^2 \rangle_S^{1/2}(ud)$ &0.85(4) fm  &  \\
 $\langle r^2 \rangle_S^{1/2}(s)$ & 1.06(9) fm  &  \\
 $\Delta\sg$ & 6.61(59) MeV  &  15.2(4) MeV \cite{gls91}  \\
 $y$ & 0.36(3) & 0.2 -- 0.3 \cite{gls91}  \\
 $\sg$ & 49.7(2.6) MeV & 45 MeV \cite{gls91} \\
  $\sigma_{KN}$ & 362(13) MeV & 395 MeV \cite{lbm95}  \\
 \hline
 \end{tabular}
 \end{table}
 
Clearly, calculations on larger lattices and smaller quark masses
will be needed to bring the systematic errors under control and
obtain a completely consistent solution on
$\Delta \sg, y$, and $\sg$ .
Eventually
dynamical fermions need to be included to complete the picture.
Nevertheless, based on what we have on a qualitative and
semi-quantitative level, we find that a consistent solution might
be close to $\Delta \sg = 6.6(6) $MeV, $y = 0.36(3)$, and
$\sg \sim 53$ MeV which are significantly different from the
present phenomenological values. We should stress that our results
on $F_S$ and $D_S$, like their counterparts in the axial couplings,
agree well with those deduced from the SU(3) breaking
pattern of the octect baryon masses and that the {\it D.I.} is
the important ingredient for this agreement. In addition, we
report the $KN \sigma$ term $\sigma_{KN} = (\hat{m} + m_s)
\langle N|\bar{u}u + \bar{d}d + 2 \bar{s}s |N\rangle/4$
in Table 1. If we assume that they are similarly depressed as in
$\sg$, we would then predict the final $\sigma_{KN}$ at
389(14) MeV. It agrees with $\sigma_{KN} = 395$ MeV from a
recent chiral analysis of $KN$ scattering \cite{lbm95}. Finally, we
note that $m_s\ss = 183(8) $MeV . Together with the kinetic
and potential
energy contribution of $ - 90 $ MeV \cite{ji95}, the strange quark
contributes about 90 MeV to the nucleon mass.

This work is partially supported by DOE
Grant DE-FG05-84ER40154. The authors wish to thank G.E. Brown,
N. Christ, T. Draper and M. Rho, for helpful comments.

{\bf Figure Captions}
 
\noindent
 
\noindent
Fig. 1 (a) Connected insertion. (b) Disconnected insertion.

\noindent
Fig. 2 (a) The lattice $g_{S, con}^L$ for the {\it C. I.}
as a function of the quark mass $ma$. The chiral limit
result is indicated by $\bullet$. (b) The form factor $g_{S,con}^L
(q^2)$ at the chiral limit.

\noindent
Fig. 3 The ratios of eq.({\ref{ratiodis}) for the scalar current are
plotted for the 3 $\kappa$ cases. ME is the fitted slope.
 
\noindent
Fig. 4 (a) The {\it D.I.} of $\ud$ and $\ss$ as a function of $ma$.
The chiral limit result is indicated by
$\bullet$. (b) The corresponding form factors.
 
\noindent
Fig. 5 The normalized form factor $g_S(q^2)/g_S(0)$ compared with
$G_E(q^2)$ and $g_A^3(q^2)$ and their respective experimental results.
 
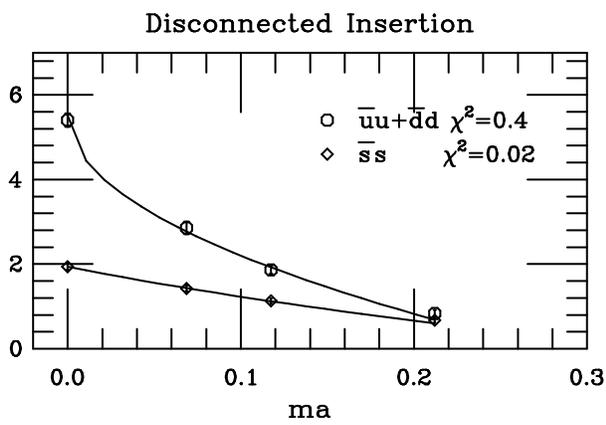
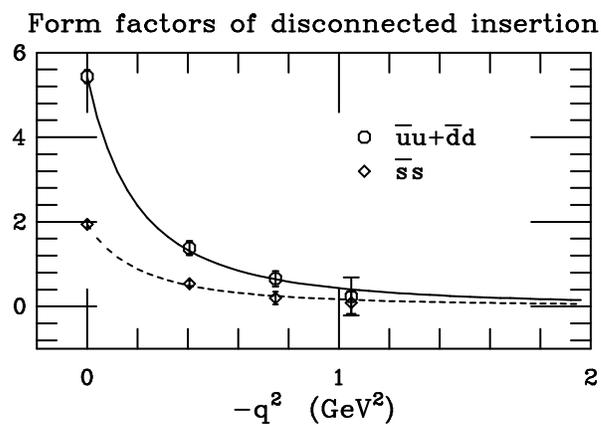
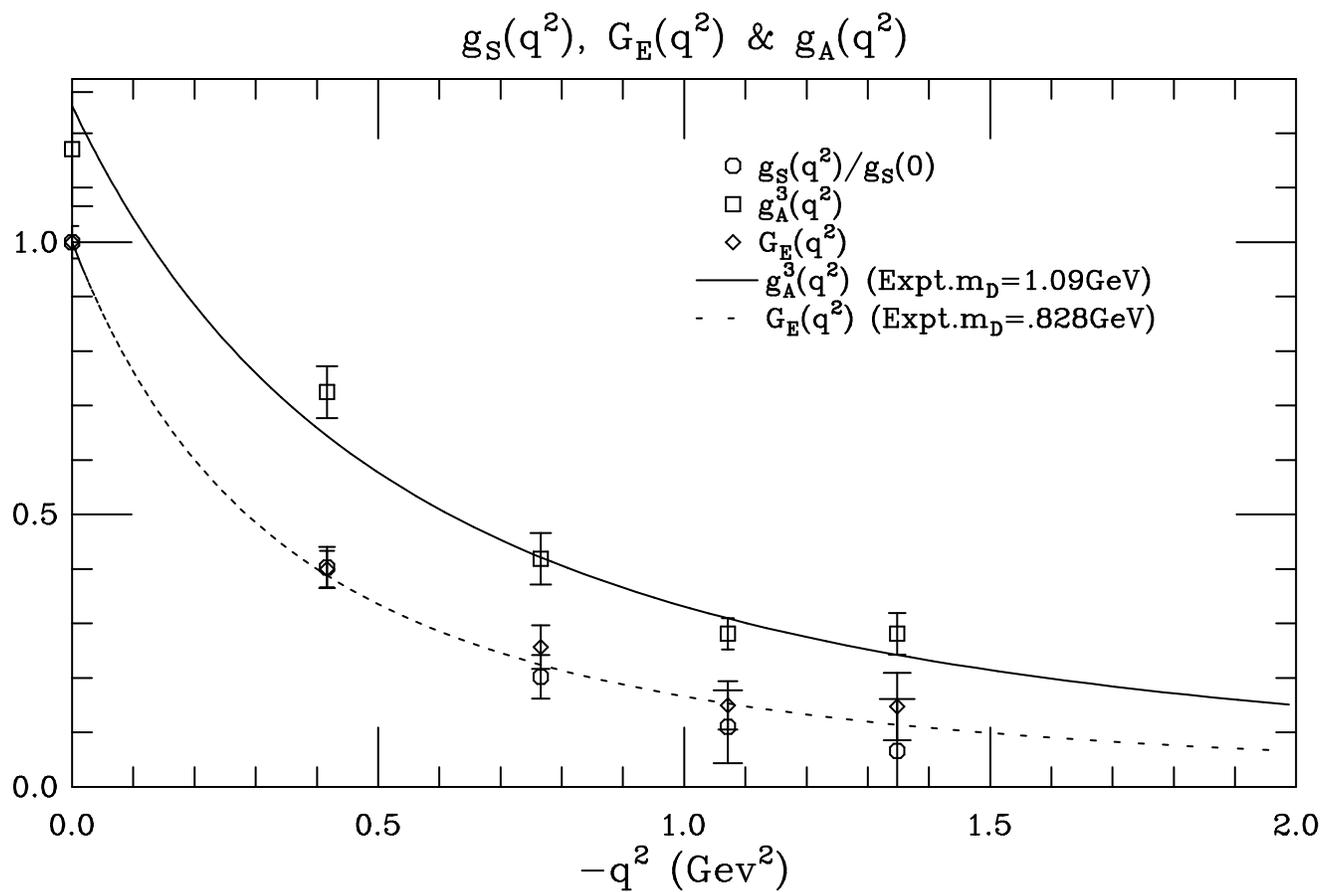
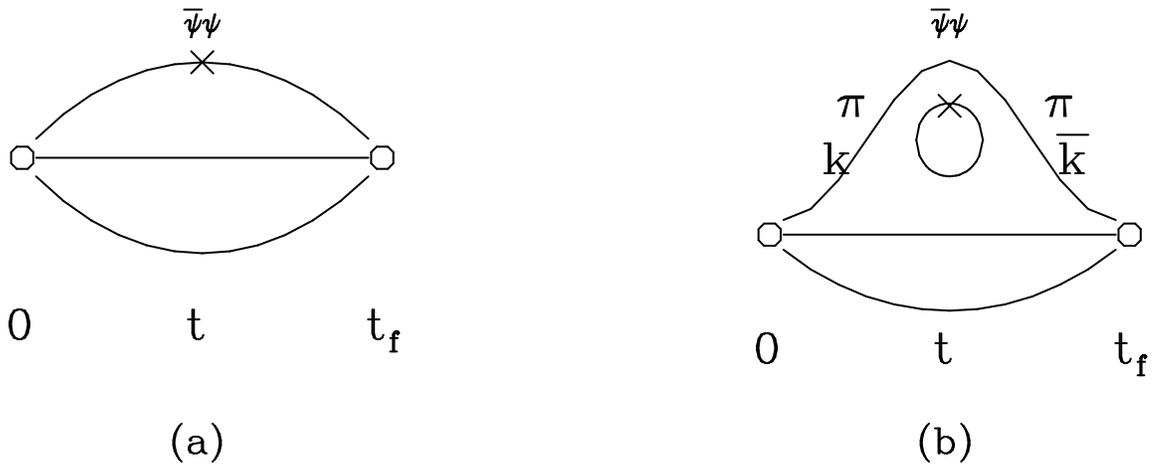
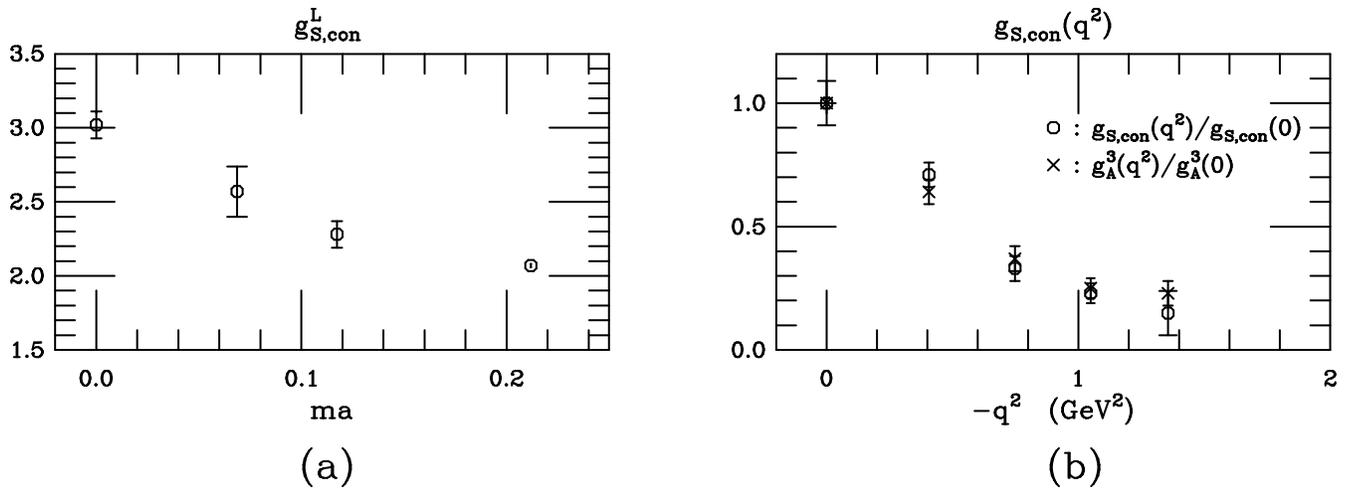
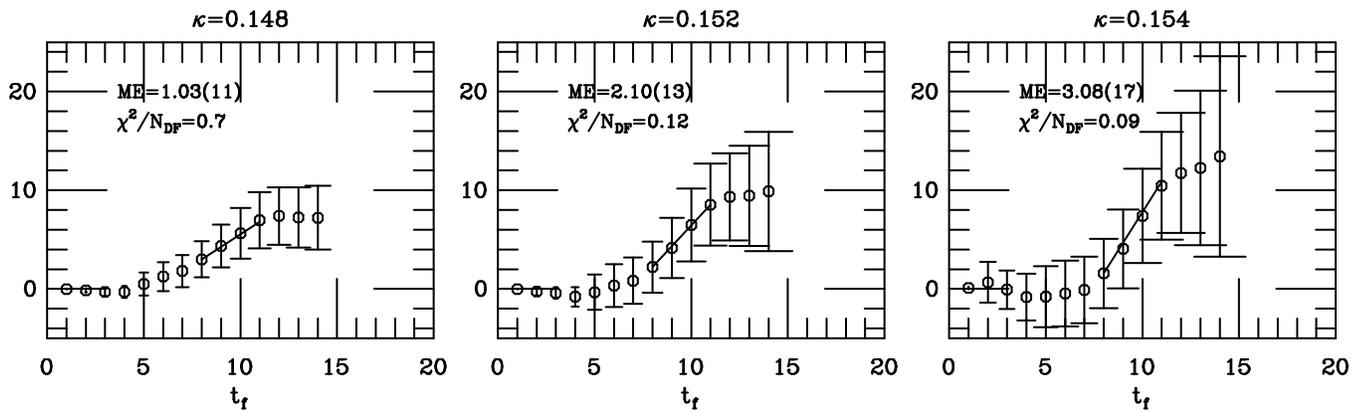
\end{document}